# A Data-driven Framework to Accelerate the Discovery of Hybrid Cathode Materials for Metal-based Batteries

Ahmed H. Biby,[1] Benjamin S. Rich,[2] and Charles B. Musgrave[*, 3,4]

[1]Materials Science and Engineering Program, University of Colorado Boulder, Boulder, CO 80309, USA
[2]Department of Chemistry, University of Colorado Boulder, Boulder, CO 80309, USA
[3]Department of Chemical Engineering, University of Utah, Salt Lake City, Utah, 84112, USA
[4]Department of Materials Science and Engineering, University of Utah, Salt Lake City, Utah, 84112, USA
*Correspondence should be addressed to charles.musgrave@utah.edu

## ABSTRACT

Selecting materials for hybrid cathodes for batteries, which involve a combination of intercalation and conversion materials, has gained interest due to their combined synergistic and compromised properties that are not attainable by their homogeneous counterparts. Herein, we present a data-driven, chemistry-agnostic, and inverse material design framework for discovering hybrid cathode materials (HCMs) for metal-based batteries. This framework systematically explores the potential materials space for any given working ion, evaluates the candidate's stability, and identifies the growth modes/adsorption of the components for a stable hybrid cathode. To demonstrate the application of the framework and its various possible outcomes, we performed a case study, for which the main design objective was to discover HCMs with an average gravimetric energy density surpassing that of the widely used high energy density NMC333 cathode material. The framework identified $LiCr_4GaS_8$-$Li_2S$ as a promising HCM that achieves an average energy density of 1,424 Wh/kg (on a lithiated cathode basis) that exceeds NMC333's maximum theoretical energy density of 1,028 Wh/kg. The identified material has several additional desirable features: 1) possession of thermodynamically stable lithiated and delithiated intercalation and conversion phases; 2) minimal volume change, upon (de)lithiation, that mitigates the high-volume change of the conversion material; 3) high energy density that ameliorates the low energy density of the intercalation material; 4) ability of the intercalation component to act as both a conductive additive and immobilizer of S, while simultaneously contributing to the total cathode energy density; 5) the intercalation material serves as an ideal support for the soft sulfur species and finally, 6) we anticipate that the life span, self-discharge, mechanical integrity, and capacity fading are better than those of conventional Li-S batteries. The developed framework was instrumental for exploring materials within the enormous potential HCM space with pre-defined battery material design objectives.

## 1. INTRODUCTION

Li-ion batteries (LIBs) are a key commercial rechargeable battery technology for a plethora of applications due to their superior performance metrics, including high energy and power densities, long cycling stability, and low maintenance. This unique combination enables practical electric vehicles, portable/mobile/wearable electronic devices, and is rapidly scaling to provide grid energy storage for renewable energy harvesting technologies.[1] However, the energy density, cost, and safety of LIBs lag the increasingly ambitious goals of every industry that employs them.[2] This has motivated the intense investigation of cathode materials for LIBs for over the past 30 years. Generally, cathode materials are classified based on their structural changes induced by their lithiation (discharging) and delithiation (charging) processes. This results in two main cathode categories, namely, intercalation/insertion materials and conversion materials. Intercalation materials undergo a topotactical phase transition upon (de)lithiation, which is a two-phase transformation that preserves the host crystal structure upon Li insertion or extraction. In contrast, conversion materials undergo a reconstructive phase transition, a two-phase transformation that results in a final phase with little or no crystallographic relationship to the initial phase.[3–6] Lithium intercalation materials are more mature than conversion materials and represent the archetypical cathode materials for LIBs. Yet, intercalation materials have nearly attained their maximum theoretical energy densities, often rely on relatively scarce and expensive elements (e.g., Co and Ni) and pose safety concerns arising from oxygen inclusion. On the other hand, conversion materials are relatively new and possess high energy densities with low materials costs, especially in the case of redox-anion chemistries. Nonetheless, they face considerable obstacles that primarily stem from their reconstructive phase transition upon (de)lithiation, which causes large volumetric changes, sluggish kinetics, and thus slow charging/discharging, and poor reversibility (rechargeability). They also tend to exhibit poor electronic and ionic conductivities.[7–9]

Several efforts have been reported that strived to design hybrid (heterostructure) cathode materials that combine both intercalation and conversion building blocks to at least achieve compromised cathode properties that are unattainable by their homogeneous counterparts. Heterostructure electrodes have several advantages, as described by Li et al.[10] First, heterostructure electrodes integrate the advantages of various components, thereby mitigating the inherent shortcomings of the standalone component materials. Second, the use of heterostructures is an effective approach for enhancing electronic conductivity via the modification of the energy band structures that narrows the band gap. Third, the introduction of an interfacial electric field between heterostructure materials lowers the ion diffusion barrier and accelerates ion diffusion. Fourth, the newly formed chemical bonds, and Van der Waals and electrostatic interactions, between the conversion and intercalation materials can improve the structural and cycling stabilities. Finally, the charge redistribution within the component materials of electrode heterostructures introduces additional active sites for energy storage. Xu et al. studied combining $VS_2$ as an intercalation material with sulfur chemistry as the conversion component in conjunction with a solid-state electrolyte. They reported that $VS_2$ provided excellent $Li^+$ and electronic conductivities, contributed to the total capacity, and concluded that $VS_2$ can serve as an ideal platform to unlock the high energy density of sulfur chemistry.[11]

Xue et al. investigated a hybrid cathode of intercalative $Mo_6S_8$ Chevrel-phase and sulfur.[12] The Chevrel-phase enhanced capacity, reduced reliance on conductive additives due to high electronic conductivity, and maintained structural integrity with minimal volumetric changes during (de)lithiation. Additionally, it immobilized sulfur species, mitigating polysulfide shuttling and capacity fading. They concluded that Chevrel-phase materials provide a robust substrate for supporting sulfur species conversion. Huang et al. developed a simple model to describe the two-stage discharging–charging behavior of the intercalation and conversion components of hybrid cathodes.[13]

Despite the research efforts on hybrid cathode materials (HCMs), the field still lacks a systematic approach to evaluate and discover potential HCMs. This is partially due to the considerable challenge of searching over the large combinatorial materials space of HCMs. Fortunately, the pursuit of high-performance HCMs inherently lends itself to the approaches of data-driven materials discovery.

Herein, we establish a data-driven, chemistry-agnostic, and inverse material design framework to accelerate the discovery of HCMs for metal-based batteries. The framework systematically explores the materials space of potential hybrid cathodes for any given working ion, evaluates the stability of HCMs, and determines the growth modes/adsorption of conversion material components to identify stable HCMs. A case study was conducted to showcase the application and potential outcomes of the framework for designing HCMs. The main design objective of the case study was to discover novel HCMs with an average gravimetric energy density greater than that of the widely used high energy density NMC333 cathode material.[14] The framework identified $LiCr_4GaS_8$-$Li_2S$ as a promising hybrid cathode that achieves the design objective of having an average energy density (1,424 Wh/kg) that exceeds the maximum theoretical energy density of NMC333 (1,028 Wh/kg), both on a lithiated cathode basis. Additionally, it demonstrates the thermodynamic stability of both the lithiated and delithiated phases. In comparison to conventional Li-S batteries, this hybrid cathode heterostructure is predicted to possess high electronic conductivities and less volume changes. Further, the life span, self-discharge, mechanical integrity, and capacity fading of the discovered HCMs are anticipated to be superior to conventional Li-S batteries. Overall, the framework proved effective in achieving pre-defined battery material design objectives by evaluating potential HCMs.

## 2. METHODOLOGY

The developed framework comprises a sequence of six steps, as shown in *Figure 1a*. The data, methods, algorithms, analytical tools, and software packages exploited in this framework have been previously established and are readily accessible. Thereby, the unique contribution of this framework lies in the integration and application of these tools to accelerate the discovery of HCMs. This framework can be used for any working ion, but herein we only considered $Li^+$ as the working ion. The detailed flowchart of the framework is depicted in *Figure S1*.

### 2.1 Data acquisition

The Materials Project (MP)[15], including the MP repository and Pymatgen[16], provide core capabilities for the framework. In defining this framework, we assumed that a given lithiated phase ($Li_nAB$) represents the fully discharged state and the corresponding completely delithiated phase (AB) is the stable charged state. As shown in *Figure 1b*, two distinct materials datasets of lithiated and delithiated/decomposed phases with energies within 100 meV/atom of the convex hull are retrieved from the MP repository. Each materials dataset encompasses various materials entries, with key features including chemical formula, formation energy per atom, structure, energy above the hull, bandgap energy, and density.

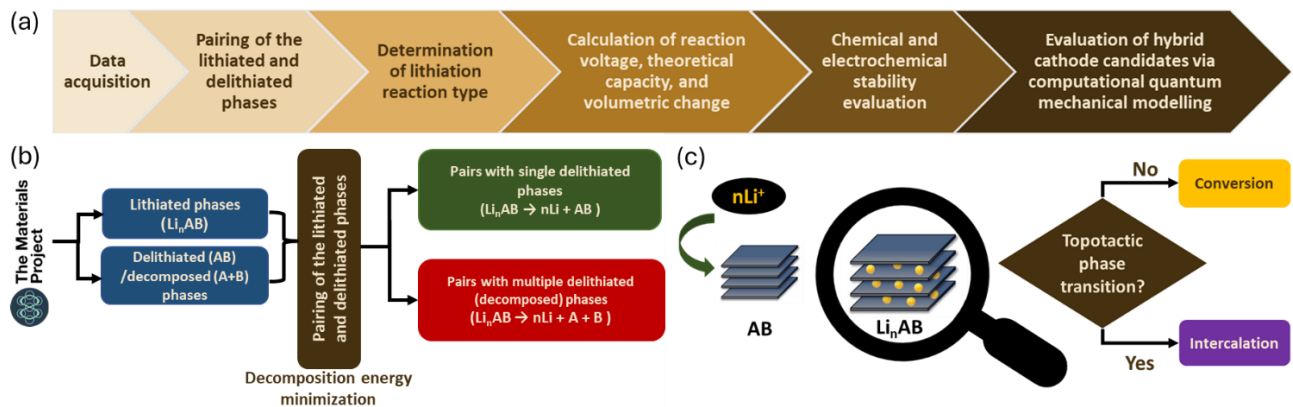

*Figure 1*: *(a) The data-driven framework scheme developed in this work for the discovery of HCMs for metal-based batteries; (b) Data acquisition and pairing of the lithiated and delithiated phases, where the datasets of lithiated and delithiated/decomposed phases are pulled from the Materials Project repository. Convex hull composition phase diagrams are constructed using SLSQP optimization to identify stable delithiated phases by minimizing the decomposition energy. Only pairs with single delithiated phases are selected for further analysis due to their better anticipated reversibility; (c) Determination of the lithiation reaction type via structural comparison between the lithiated-delithiated pair, where crystallographic similarity between lithiated and delithiated structures is assessed with Pymatgen's StructureMatcher to classify the (de)lithiation-induced structural changes as intercalation or conversion reactions.*

## 2.2 Pairing of the lithiated and delithiated phases

To thermodynamically pair the Li phases to the delithiated/decomposed phases, the Sequential Least SQuares Programming (SLSQP) optimizer from the Scipy python library[17] was exploited to construct a convex hull composition phase diagram to perform a thermodynamic stability analysis for each composition space available in the materials dataset. SLSQP was applied on the formation energies of the possible delithiated/decomposed phases to minimize the decomposition energy of a given Li phase, consequently determining the corresponding delithiated/decomposed phase(s), as shown in *Figure 1b*. The pairs with single delithiated phases are adopted for the subsequent steps, while pairs with multiple delithiated (decomposed) phases are excluded from further consideration due to their anticipated poor reversibility as components of any potential hybrid cathode upon (de)lithiation.

## 2.3 Determination of lithiation reaction type

To determine the (de)lithiation reaction type for pairs with single delithiated phases, the crystallographic similarity between the lithiated and delithiated structures is evaluated for a given pair. This crystallographic comparison determines whether the (de)lithiation reaction is a topotactic phase transition (intercalation) or a reconstructive phase transition (conversion), as shown in *Figure 1c*. The StructureMatcher module from the Pymatgen library is utilized for structural comparison. The flowchart of the StructureMatcher module is shown in *Figure S2*. In some cases, a single lithiated phase ($Li_nAB$) has several possible single delithiated polymorphs that share the same chemical formula (AB), yet that have different structures. Accordingly, crystallographic comparisons for a given lithiated phase can identify multiple possible competing lithiation reactions. Thus, two cases arise; either all possible reactions are 1) of the same lithiation reaction type, or 2) a mix of the two lithiation reaction types. In the former case, the lithiated phase is paired to the delithiated phase that exhibits the lowest formation energy, thereby allowing direct identification of the lithiation reaction. In the latter case, the lithiated phase is paired to the delithiated phase with the lowest formation energy within the intercalation lithiation type, as the formation energies of all potential delithiated polymorphs are closely matched and mostly fall within the range of DFT error. Moreover, the thermodynamic feasibility of a lithiation reaction ($Li^+$ intercalation) is substantiated by the comparatively lower kinetic barriers associated with topotactical intercalation reactions, which induce marginal crystallographic changes. In contrast, reconstructive conversion reactions induce significant crystallographic changes. It is pertinent to note that from this step onwards, the phases under consideration possess energies ≤ 50 meV/atom above the hull.

## 2.4 Calculation of reaction voltage, theoretical capacity, and volumetric change

For each lithiated-delithiated (Li-deLi) pair, key cathode metrics are computed using the materials' features obtained from the MP to guide subsequent analyses and procedures. The reaction voltage, theoretical capacities, and volumetric change are evaluated using *Equations 1-5*, *respectively*.[17,18]

$$\Delta G_r \approx E_{DFT}(Li_n A) - E_{DFT}(Li_{n-x} A) - x\, E_{DFT}(Li) \tag{1}$$

$$V = -\frac{\Delta G_r}{x\, z\, F} \tag{2}$$

$$\varepsilon_m = \frac{\Delta G_r}{mass_{Li\ phase}} \tag{3}$$

$$\varepsilon_v = \frac{\Delta G_r}{volume_{Li\ phase}} \tag{4}$$

$$\Delta volume = \frac{|volume_{Li\ phase} - volume_{deLi\ phase}|}{volume_{Li\ phase}} \tag{5}$$

Where $\Delta G_r$ is the reaction Gibbs free energy, $E_{DFT}$ represents the energies of $Li_nA$, $Li_{n-x}A$, and Li that are calculated from DFT-computed total energies, $x$ is the number of $Li^+$ ions involved in the lithiation reaction, V is the reaction voltage, $z$ is the charge on the working-ion, which is 1 in case of $Li^+$), $F$ is Faraday's constant, $\varepsilon_m$ is the gravimetric energy density, and $\varepsilon_v$ is the volumetric energy density.

## 2.5 Chemical and electrochemical stability evaluation

Herein, the computational scheme to evaluate the chemical and electrochemical stabilities of heterogeneous solid interfaces is adopted from Zhu et al.[20] However, we introduced criterion 2, as shown in *Figure 2,* to account for the energy density contribution requirement for both components of the HCM. For hybrid cathode candidates that successfully passed previous down selection steps, the intercalation (intcl) and conversion (conv) component materials must have compatible chemical and electrochemical windows, as illustrated in *Figure 2*. There are four criteria for the chemical and electrochemical stabilities of the two cathode building blocks:

> **Criterion 1:** The intrinsic and interfacial electrochemical stability windows of the component materials against the working ion (Li$^+$) must overlap to ensure that both building blocks along with their interface coexist over the same voltage range ($\Delta\Phi$) without decomposition. This criterion is evaluated using *Equations 6-9*.
>
> **Criterion 2:** The reaction voltages of the intercalation and conversion component materials must lie within the overlap determined in *Criterion 1* to ensure that both component materials can be charged/discharged without decomposing. This criterion is evaluated using *Equations 2* and *6-9*.
>
> **Criterion 3:** For each non-Li element within the combined compositional space of component materials, the intrinsic chemical stability windows of the building blocks against the non-Li element must overlap to ensure that a common chemical potential value ($\mu_{non-Li}$) can be achieved across the interface to avoid decomposition/cross-diffusion. This criterion is evaluated using *Equation 10*.
>
> **Criterion 4:** The interface must be chemically stable against decomposition to ensure the chemical stability of the interface under conditions of no applied voltage. This criterion also ensures material stability during cell preparation and is evaluated using *Equations 6-7* and *11*.

For the interfacial Criteria 1 and 4, C$_{interface}$, C$_{intcl}$, C$_{conv}$, E$_{interface}$, E$_{intcl}$, and E$_{conv}$ represent the compositions and energies of the pseudobinary interface, intercalation and conversion materials, as shown in *Equation 6* and *7*.

$$C_{interface}(C_{intcl}, C_{conv}, x) = xC_{intcl} + (1-x)C_{conv} \tag{6}$$

$$E_{interface}(E_{intcl}, E_{conv}, x) = xE_{intcl} + (1-x)E_{conv} \tag{7}$$

where $x$ is assumed to be 0.5; because it fairly provides the minimum mutual reaction energies, as results suggest in Zhu et al.. [20]

The grand potential phase diagrams at $T = 300$ K[21] are constructed using the Pymatgen library and the MP database to identify the competing phase equilibria (eq) and to calculate the decomposition energies and stability windows of the intercalation and conversion phases using the following equations:

$$\Delta E_D^{open}(phase, \mu_{Li}) = E_{eq}[C_{eq}(C, \mu_{Li})] - E(phase) - \Delta n_{Li}\mu_{Li} \tag{8}$$

$$\Delta E_D^{open}(intcl, conv, \mu_{Li}) = E_{eq}[C_{eq}(C_{interface}, \mu_{Li})] - E_{interface}(intcl, conv) - \Delta n_{Li}\mu_{Li} \tag{9}$$

$$\Delta E_D^{open}(phase, \mu_{non-Li}) = E_{eq}[C_{eq}(C, \mu_{non-Li})] - E(phase) - \Delta n_{non-Li}\mu_{non-Li} \tag{10}$$

$$\Delta E_D(intcl, conv) = E_{eq}(C_{interface}) - E_{interface}(intcl, conv) \tag{11}$$

where $\Delta E_D^{open}$ is the decomposition reaction energy at a given chemical potential $\mu$, $C_{eq}$ and $E_{eq}$ are the composition and the total formation energy of the competing phase equilibria at a given chemical potential $\mu$, $E(phase)$ is the formation energy of the phase of interest, and $\Delta n$ is the change in the number atoms of a given element at chemical potential $\mu$.

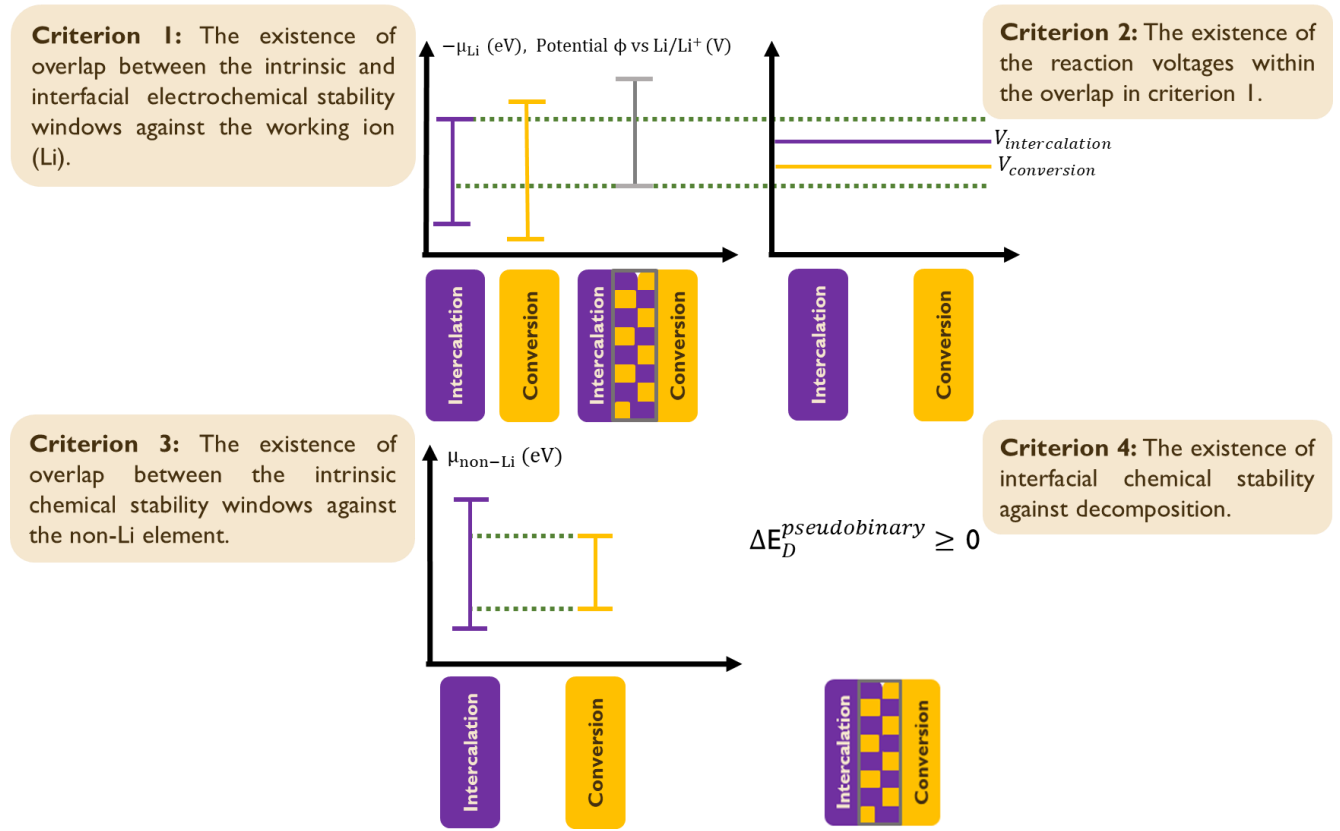

*Figure 2: Chemical and electrochemical stability evaluation scheme. The stability of hybrid cathode candidates is assessed based on four criteria: **Criterion 1** ensures overlapping electrochemical stability windows of intercalation and conversion materials to prevent decomposition (Equations 6-9); **Criterion 2** requires reaction voltages within the overlap of*

*Criterion 1 (Equations 2 and 6-9);* **Criterion 3** *mandates overlapping chemical stability windows for non-Li elements to avoid decomposition/cross-diffusion (Equation 10).* **Criterion 4** *ensures interface stability in the absence of an applied voltage (Equations 6, 7 and 11).*

In this study, the stable chemical and electrochemical windows were defined as the potential ranges over which the phase of interest demonstrates decomposition energies ≥ -50 meV/atom. This threshold is selected to account for the DFT margin of error that may introduce errors into the calculated thermodynamic formation energies.

Relying solely on thermodynamic criteria to screen materials for various applications can be stringent and often prohibitively difficult to satisfy. This is especially true for grand canonical composition phase diagrams that include numerous elemental compositions. To illustrate this, the intrinsic electrochemical stability windows against $Li^+$ (working ion) are calculated based on the stability of the decomposition products. Accordingly, these calculations usually underestimate the experimental electrochemical stability windows due to their inability to account for the kinetics that arise from the activation barriers to form the decomposition products. That is, they fail to identify materials that are sufficiently metastable because they are kinetically trapped from forming decomposition products at the conditions of interest. To remedy this, the electrochemical stability windows can be computed based on indirect decomposition via (de)lithiation of the cathode material, as explained in Schwietert et al. and Zhu et al..[22,23] Unfortunately, this approach requires extensive DFT calculations, which renders it impractical for high-throughput data-driven materials screening studies. Alternatively, reasonable tolerances can be considered to widen the electrochemical stability windows for Criteria 1 and 2. For Criterion 3, if there are reasonable gaps between the chemical windows, transition-state theory (TST) calculations can be used to predict whether the material is kinetically trapped by estimating the diffusivity of the thermodynamically unstable species. This allows one to assess the probability that these species diffuse out the cathode material to cause decomposition to occur at a given temperature. These species are usually less mobile than Li, increasing the likelihood that these materials are sufficiently metastable to persist sufficiently for a given application.

## 2.6 Evaluation of hybrid cathode candidates via computational quantum mechanical modelling

For stable hybrid cathode candidates that pass previous screening steps, computational quantum mechanical modelling is exploited to study the nature of the lithiated and delithiated hybrid cathode interfaces. For the case of two crystalline solid cathode materials, the dominant interfacial growth modes can be identified to offer insights in cases where interface modifications, such as applying surfactants, are needed. For the case of a crystalline solid cathode and a molecular species (e.g., $S_8$), the thermodynamic favorability of adsorption is assessed to provide insights into the immobilization of the molecular species on the crystalline solid cathode surface, for example to evaluate cathode stability against polysulfide shuttling.

### 2.6.1 Interfacial growth mode

For a given hybrid cathode, low-index facets of the intercalation and conversion components that minimize the interfacial von Mises strain, with a maximum limit of 5%,[24] should be chosen. DFT calculations can then be utilized to compute the interfacial ($E_{interface}$), surface ($E(slab)$), and adhesion ($W_{ad}$) energies using the following equations:[18,25]

$$\text{Surface energy} \quad \gamma = \frac{1}{2A}[E(slab) - N \cdot E(bulk)] \quad (12)$$

$$\text{Adhesion energy} \quad W_{ad} = \frac{1}{A}[E_{substrate} + E_{film} - E_{interface}] \quad (13)$$

$$\text{Interface energy} \quad \sigma_{interface} = \frac{1}{A}[\gamma_{substrate} + \gamma_{film} - W_{ad}] \quad (14)$$

Where $A$ is the surface area of a surface/interface, $E(slab)$ is the total energy of a given facet, $E(bulk)$ is the bulk total energy/atom of a given facet of the phase of interest, $N$ is the number of atoms in the surface slab, and $E_{substrate}$, $E_{film}$, and $E_{interface}$ are the total energies of the substrate, film, and interface, respectively.

The relation between $\gamma_{film}$, $\gamma_{substrate}$, and $\sigma_{interface}$ defines the dominant interfacial growth mode. Thus, if $\gamma_{film} \leq \gamma_{substrate} - \sigma_{interface}$, the dominant growth mode will be layer-by-layer (Frank–Van der Merwe) growth. Otherwise, the dominant interfacial growth is island (Volmer-Weber) growth.[26]

### 2.6.2 Adsorption study

An adsorption study can be conducted for a given hybrid cathode if one of its components is a molecular species. Surface energies at neutral charge (unbiased) conditions for the low-index facets are evaluated using DFT to identify the facet with the lowest surface energy (most abundant facet) for the adsorbent. Binding energies can then be computed for adsorption at different sites on the most stable facet. To make binding energies representative of working conditions, all DFT calculations are evaluated within the grand canonical ensemble, where the electron reservoir potential is sampled over the operating potential window of the cathode.

### 3. RESULTS AND DISCUSSION

The primary motivations for applying the framework in this work are 1) to provide an illustrative case study that demonstrates its application and potential outcomes and 2) to achieve the design objective of discovering novel HCMs with an average gravimetric energy density that surpasses that of NMC333, which is a widely-used intercalation cathode material that boasts a maximum theoretical gravimetric energy density of 1,028 Wh/kg, based on the mass of the lithiated cathode.[14] The detailed flowchart for designing a high-energy density HCM is illustrated in **Figure S3**.

Two distinct datasets containing lithiated and delithiated/decomposed phases were retrieved from the MP repository. These datasets were filtered to include only phases with energies within 100 meV/atom of the convex hull, while phases containing elements that are scarce, toxic, or radioactive were excluded as detailed in **Table S1**. Also, the phases that include fluorine or hydroxyl groups were excluded due to their limited reversibility and propensity to trigger parasitic reactions, respectively. The lithiated phases dataset comprised 11,688 phases. The pairing of lithiated and delithiated phases resulted in 1,909 pairs with single delithiated phases and 9,759 pairs with multiple delithiated (decomposed) phases. The structural comparison, via the StructureMatcher module from the Pymatgen library, yielded 412 intercalation phases and 1,497 conversion phases. Our evaluation of the calculated theoretical capacities led to our conclusion that the only way to surpass NMC333's gravimetric energy density of 1,028 Wh/kg is by incorporating sulfur as the conversion component material. The chemical and electrochemical stability evaluation scheme was exclusively applied to the 412 HCMs. Each HCM was composed of one of the 412 intercalation building blocks combined with $Li_2S$ as the conversion building block. Interestingly, applying the complete thermodynamic stability criteria, stated in the chemical and electrochemical stability evaluation scheme, did not result in any of the 412 HCMs passing, especially for Criterion 2. This is attributed to the multiple oxidation states, high polarization and reactivity of sulfur. Nonetheless, Schwietert et al. showed that the estimation of the thermodynamic formation energy of the decomposed products frequently underestimates the actual electrochemical stability range.[23] Hence, a less conservative version of Criterion 2 was adopted to account for the overpotentials of the de(lithiation) reaction, which is an inner-sphere electron transfer redox reaction involving bonds breaking/formation that typically requires activation energy more than 0.4 eV.[27] Thus, a margin of 0.5 eV was added to both limits of the overlap window of Criterion 2. This modified criterion led to the identification of 12 metastable lithiated intercalation-$Li_2S$ HCMs. Also, the chemical stability evaluation scheme based on Criteria 3 and 4, as Criteria 1 and 2 are applicable only to the lithiated phases, were applied to the corresponding 12 delithiated intercalation-$S_8$ HCMs, which resulted in only 1 stable candidate ($LiCr_4GaS_8$-$Li_2S$), as depicted in *Figures 3a* and *3b* and tabulated in *Table S2*. *Figure 4* summarizes the results of the framework up to step 5.

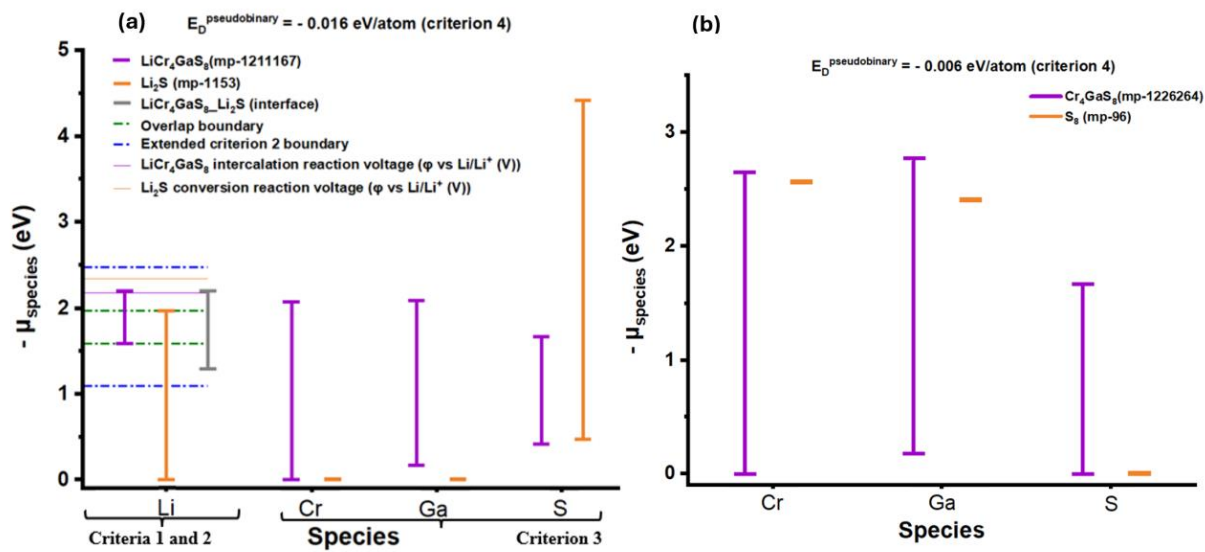

*Figure 3*: *(a) Chemical and electrochemical stability evaluation scheme for each species in $Li_2S$ and $LiCr_4GaS_8$ (lithiated phases), where the hybrid material passed Criterion 1 by having*

*overlapping electrochemical stability windows of intercalation and conversion components, passed the modified version of Criterion 2 by having straddled reaction voltages within the extended Criterion 2 boundaries, passed Criterion 3 by having overlapping chemical stability windows for non-Li elements, and passed Criterion 4 by having $E_D^{pseudobinary} \geq$ -50 meV/atom. (b) Chemical stability evaluation scheme (Criteria 3 and 4) for each species in $S_8$ and $Cr_4GaS_8$ (delithiated hybrid components), where Criteria 1 and 2 are irrelevant here because the intercalation and conversion components are free from Li ion (the working ion). The delithiated hybrid components passed Criterion 3 by having overlapping chemical stability windows for non-Li elements, and passed Criterion 4 by having $E_D^{pseudobinary} \geq$ -50 meV/atom.*

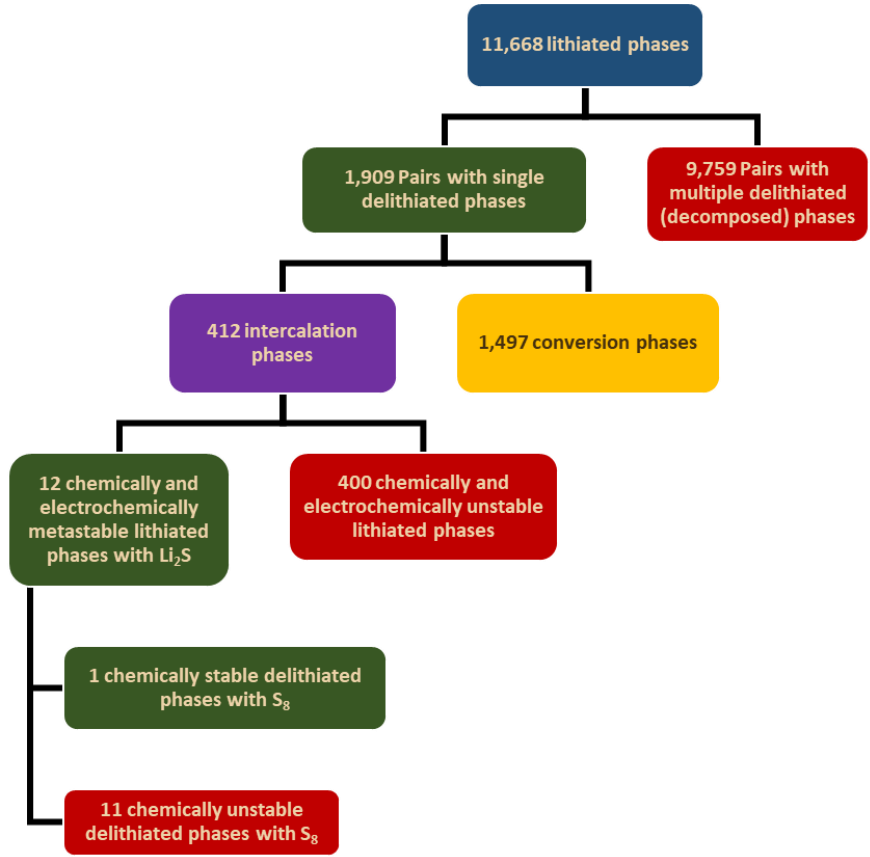

*Figure 4: The summary of the outcomes of the framework case study for designing high-energy density HCMs. A dataset of 11,688 lithiated phases was pulled from the MP repository. This resulted in 1,909 single-phase and 9,759 multi-phase pairs. Classification categorized 412 materials as intercalation phases and 1,497 materials as conversion phases. From our evaluation, we concluded that surpassing NMC333's energy density requires sulfur as the conversion component. Stability evaluation on 412 HCMs (intercalation phase + $Li_2S$) led to no candidates passing Criterion 2. A modified Criterion 2, with a 0.5 V margin, identified 12 metastable HCM. Chemical stability evaluation on the delithiated HCMs reduced the candidate pairs to one stable option: $LiCr_4GaS_8$-$Li_2S$.*

### 3.1 Examination of the identified hybrid cathode candidate by DFT

Upon being identified by our framework as the most stable candidate, the LiCr$_4$GaS$_8$-Li$_2$S (mp-1211167-mp-1153) HCM was further evaluated using DFT. Specifically, further investigation was performed to confirm whether Ga is kinetically stabilized in LiCr$_4$GaS$_8$, given the slight gap (0.168 eV) between the μ$_{Ga}$ windows of LiCr$_4$GaS$_8$ and Li$_2$S, as depicted in *Figure 3a*. Additionally, this study identified the dominant interfacial growth mode of Li$_2$S on LiCr$_4$GaS$_8$ and the adsorption energetics of S$_8$ molecules on Cr$_4$GaS$_8$ to evaluate the stabilization effect of Cr$_4$GaS$_8$ *for S$_8$*.

### 3.1.1 Ga migration barrier calculation

To determine the stability of LiCr$_4$GaS$_8$ against decomposition, we computed the energy barrier to Ga diffusion using the climbing-image solid-state nudged elastic band (CI SS-NEB) method.[28,29] The computational details are provided in the Computational details. Ga resides on only one unique site within LiCr$_4$GaS$_8$, shown in *Figure 5a*, and the minimum energy pathway (MEP) for Ga migration between adjacent Ga sites was computed. The NEB-computed results predict that the migration energy barrier for Ga in LiCr$_4$GaS$_8$ is 3.32 eV using the PBE functional. Ga diffusion along the MEP results in a transition state that occurs where Ga passes through an S triangular plane (bottleneck), as shown in *Figure 5a*. These findings agree with the fact that the space available within the sulfur triangular plane (0.57 Å$^2$) constitutes the narrowest point along the Ga MEP, which is smaller than the Ga$^{3+}$ cross-sectional area (0.69 Å$^2$). We note that the exponential factor for the rate of diffusion of Ga in LiCr$_4$GaS$_8$ at 298 K is ~50 orders of magnitude smaller than that of Li$^+$ in some ceramic electrolytes[30], indicating that Ga is kinetically-stabilized in LiCr$_4$GaS$_8$.

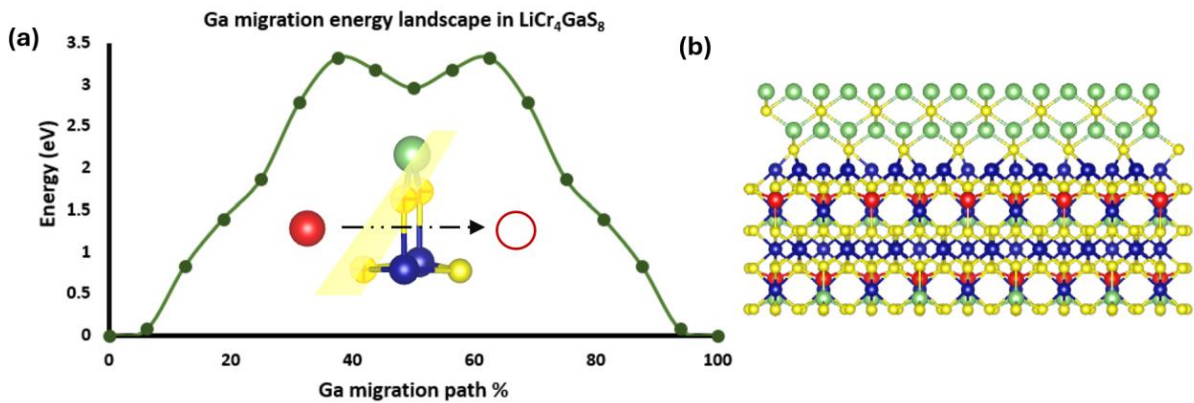

*Figure 5*: *(a) The Ga (red) migration path in LiCr$_4$GaS$_8$. The yellow plane passes through the plane defined by the 3 sulfur atoms (yellow), which indicates the position of the migration barrier. The green, blue, hollow red spheres represent Li, Cr, and Ga symmetric site, respectively. The area available within the sulfur triangular plane (0.57 Å$^2$) constitutes the narrowest point along the Ga migration pathway, which is smaller than the Ga$^{3+}$ cross-sectional area of 0.69 Å$^2$. The NEB-computed Ga diffusion minimum energy path in LiCr$_4$GaS$_8$, where the migration energy barrier for Ga occurs at the plane defined by the 3 sulfur atoms for LiCr$_4$GaS$_8$ and is computed to be 3.32 eV. (b) The interface formed between LiCr$_4$GaS$_8$ (111) and Li$_2$S (110) possesses the lowest von Mises strain of 1% and was adopted for the interfacial growth mode study. The green, yellow, blue, and red spheres represent Li, S, Cr, and Ga atoms, respectively.*

### 3.1.2 The interfacial growth mode of Li$_2$S on LiCr$_4$GaS$_8$

The interfacial growth mode of Li$_2$S on LiCr$_4$GaS$_8$ was evaluated as described in Section 2.6.1 and in the Computational details. The interface between LiCr$_4$GaS$_8$ (111) and Li$_2$S (110) achieved the lowest von Mises strain of 1% and was adopted to perform the interfacial growth mode analysis, as shown in *Figure 5b*. The results show that the dominant interfacial growth mode of Li$_2$S on LiCr$_4$GaS$_8$ is the Volmer-Weber island growth mode because $\gamma_{Li2S} > \gamma_{LiCr4GaS8} - \sigma_{interface}$.

### 3.1.3 The adsorption of S$_8$ on Cr$_4$GaS$_8$

Although sulfur can form solid crystals, these crystals are considered weak molecular solids held together by relatively weak Van der Waals forces. Therefore, understanding the adsorption of S$_8$ on Cr$_4$GaS$_8$ is a critical first step for predicting the interfacial growth of sulfur on Cr$_4$GaS$_8$. To this end, we evaluated the adsorption of S$_8$ on Cr$_4$GaS$_8$ as described in Section 2.6.2 and the computational details. Surface energies for the (100), (110), and (111) facets of Cr$_4$GaS$_8$ were compared at neutral charge, as shown in *Figure S4*. The (100) facet possesses the lowest surface energy and was adopted for the adsorption study, as shown in *Figure 6*. The electron reservoir potential was sampled over the charging (delithiation) operating window at 2 V, 3.5 V and 5 V (Li/Li$^+$), which spans the range over which S$_8$ exists (above the oxidation onset of Li$_2$S). The results revealed favorable adsorption energies of S$_8$ on the S-S bridge site of Cr$_4$GaS$_8$ between -0.65 and -0.85 eV, suggesting that Cr$_4$GaS$_8$ stabilizes and immobilizes S$_8$ within the studied potential region, as shown in *Figure 6*.

### 3.2 Features of the discovered HCM

LiCr$_4$GaS$_8$-Li$_2$S, which we discovered using our framework is predicted to achieve the design objective of an average energy density of 1,424 Wh/kg on a lithiated cathode basis, which exceeds the maximum theoretical energy density of NMC333 of 1,028 Wh/kg. Furthermore, LiCr$_4$GaS$_8$-Li$_2$S provides several advantages. First, according to MP, both the lithiated and delithiated intercalation and conversion phases of LiCr$_4$GaS$_8$-Li$_2$S lie on the hull, meaning that they are thermodynamically stable. Second, according to MP, both the lithiated and delithiated intercalation phases of LiCr$_4$GaS$_8$-Li$_2$S have 0 eV bandgap, which relieves the low electronic conductivities of the conversion building block that results from the wide bandgaps of the lithiated (3.54 eV) and delithiated (2.51 eV) conversion phases of LiCr$_4$GaS$_8$-Li$_2$S. Third, the volume change of LiCr$_4$GaS$_8$ (the intercalation component) upon (de)lithiation is only 7%, which mitigates the high-volume change (80%) of Li$_2$S (the conversion component). Fourth, the Li$_2$S (the conversion component) possesses a high energy density of 2,739 Wh/kg, which alleviates the low energy density of the intercalation material of only 108 Wh/kg. Fifth, LiCr$_4$GaS$_8$ (the intercalation component) can act as both a conductive additive and immobilizer of S to mitigate sulfur species shuttling while actively contributing to the energy density of the cathode. Sixth, the intercalation material serves as an ideal substrate for supporting the molecular sulfur species. Finally, the life span, self-discharge, mechanical integrity, and capacity fading of batteries with this hybrid cathode are anticipated to be improved over conventional Li-S batteries.

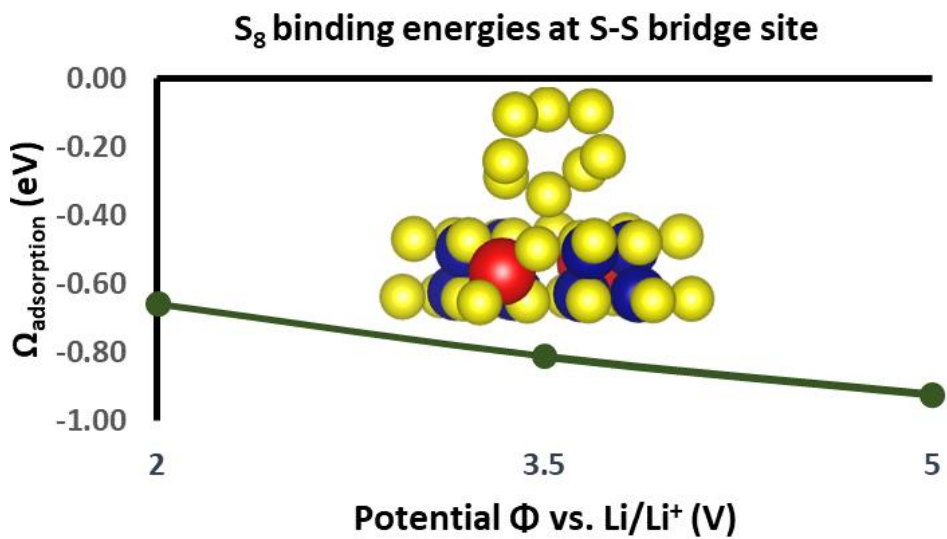

*Figure 6*: *The adsorbed $S_8$ species on the (100) facet of $Cr_4GaS_8$. The (100) facet has the lowest surface energy and was adopted for the adsorption study. The yellow, blue, and red spheres represent S, Cr, and Ga atoms, respectively. $S_8$ binding energies at the S-S bridge site of $Cr_4GaS_8$, where the electron reservoir potential was sampled over the charging (delithiation) operating window. The results revealed that $Cr_4GaS_8$ stabilizes and immobilizes $S_8$.*

## CONCLUSIONS

This study serves as a blueprint for systematic data-driven design of HCMs for metal-based batteries that can be customized to specific design objectives. An illustrative case study was performed to demonstrate the framework's application and its potential outcomes. The main design objective was to discover novel HCMs with an average gravimetric energy density surpassing that of NMC333. The framework identified $LiCr_4GaS_8$-$Li_2S$ (mp-1211167-mp-1153) as a promising HCM that achieves an average energy density of 1,424 Wh/kg, exceeding the maximum theoretical energy density of NMC333 (1,028 Wh/kg).

The comprehensiveness and the fidelity of this case study is limited by the available phases and the level of theory used to predict the properties of the materials in the MP repository. Interestingly, the electrochemical and chemical stability scheme proved to be the bottleneck of the framework due to its intricate thermodynamic criteria. This is especially true for composition phase diagrams with the extensive elemental compositions of the HCMs and under the grand canonical ensemble conditions of battery environments. We observe that finding stable interfaces is a challenging task for materials spaces including sulfur because it possesses multiple oxidation states, high polarization and reactivity.

We suggest that experimental validation of the identified HCM, $LiCr_4GaS_8$-$Li_2S$, is needed. Moreover, a multi-objective optimization study is crucial to identifying the critical mixing ratio of the intercalation and conversion component materials that maximizes both the rate capability and the energy density of the HCM. Additionally, the properties of the HCM can be further refined through compositional modifications such as substituting Ga with B or Al and exploiting defect engineering techniques. Further avenues of exploration could involve investigating and maximizing the HCM lithiation capacity. Finally, the developed framework can be systematically exploited to explore the potential HCM space for any given working ion and any chemistry of interest with pre-defined battery material design objectives.

# COMPUTATIONAL DETAILS

For Ga migration barrier and interfacial growth mode calculations, DFT calculations were performed using the Vienna ab initio simulation package (VASP)[31] with the Perdew, Burke and Ernzerhof (PBE) generalized gradient approximation (GGA)[32] exchange-correlation (XC) functional and the projector augmented-wave (PAW) method.[33] A plane wave cutoff energy of 520 eV was applied.

**Ga migration barrier calculation**

To compute the kinetic stability of Ga in $Cr_4GaS_{48}$, a Γ-point centered 2×2×2 k-point mesh was used for Brillouin zone integration. The structural relaxations were completed when the residual force on each atom was < 0.01 eV/Å. The bulk structure of $Cr_4GaS_{48}$ (mp-1226264) was pulled from the Materials Project repository. Ga migration in $Cr_4GaS_4$ occurs via a vacancy hopping mechanism. The Ga ion passes through the bottleneck of the "critical triangle" formed between three S atoms. A 2×2×2 supercell that has a Ga defect was created. Minimum energy path (MEP) calculations between two optimized configurations each with a Ga vacancy were completed using the climbing image solid-state nudged elastic band (CI SS-NEB) method[28,29,34] where the migration path is divided into a number of equidistant configurations (images). This study used seven images, each representing a step along the MEP from the initial state to the transition state because the MEP is symmetric. These images are linked by springs and their positions were optimized. The optimization process used in the NEB method employs both the spring force, which acts along the migration path to prevent the images from settling into local minima, and the true force perpendicular to the migration path to ensure the discovery of the lowest energy pathway.

**The interfacial growth mode of $Li_2S$ on $LiCr_4GaS_8$**

To optimize heterostructures, the relaxation of electronic degrees of freedom stops when both the change in free energy and the change in band structure energy between two steps are < $10^{-5}$ eV. Pymatgen was employed to automatically create a Γ-point centered k-point mesh for Brillouin zone integration with grid density of 1200 points per the total number of atoms of a given structure.[16] The bulk structures of $LiCr_4GaS_8$ (mp-1211167) and $Li_2S$ (mp-1153) were pulled from the Materials Project repository. Also, Pymatgen was used to cut the surface facets, calculate the von Mises strains, and construct the interfaces. An interface formed between $LiCr_4GaS_8$ (111) and $Li_2S$ (110) achieved the lowest von Mises strain and was adopted for the interfacial growth mode study. The interface formed between $LiCr_4GaS_8$ (111) and $Li_2S$ (110) was constructed out of one layer of each component with 1.5 Å of interfacial spacing and vacuum slab of 25 Å to remove spurious effects from interactions with periodic images. Selective dynamics was used to freeze all the atoms except for the interfacial frontier atoms.

**The adsorption of $S_8$ on $Cr_4GaS_8$**

All adsorption calculations were evaluated using the JDFTx program[35] with the PBE functional with dispersion interactions corrected with the Grimme semi-empirical D3 dispersion correction.[36] Core electrons were represented using the GBRV pseudopotential set.[37] The starting geometry for bulk phase of material was taken from MP. A planewave energy cutoff of 544 eV was compared to a convergence analysis in the range of 490-600 eV and was shown to be near convergence.

For optimization of the bulk phase, 5×5×5 k-point mesh was chosen as a balance between computational efficiency and accuracy based on a convergence test. For surface calculations, surfaces were cut using Pymatgen, and a standard k-point density of 24/atom was used for assigning k-point meshes. Solvation of surfaces was modeled implicitly using the GLSSA13 linear PCM model[38], where glyme was modeled using 0.5 M $Na^+$ and 0.5 M $F^-$ as the electrolyte. Surface energies for the (100), (110), and (111) surface facets were compared at neutral charge (unbiased) conditions. Alternative terminations to these surface facets were constructed to explore possible lower energy configurations. As shown in *Figure S4*, the (100) surface facet was found to be the most stable surface, and thus was chosen for investigating sulphur binding energies.

Binding energies for sulphur were computed within the grand canonical ensemble with the electron reservoir potential sampled at +2.00 V, +3.50 V, and +5.00 V wrt $Li/Li^+$. Sulphur binding was investigated by direct adsorption of $S_8$. Regardless of the binding site, $S_8$ prefers binding at oxidative potentials. This likely indicates a positive correlation between $S_8$ binding energy to positive surface charge. Given the relatively high potential of zero charge (PZC) of the (100) surface facet (3.61 V wrt $Li/Li^+$), other surfaces that may exist with lower surface areas would likely bind $S_8$ stronger at equivalent biases. At +2.00 V, +3.50 V, and +5.00 V wrt $Li/Li^+$, $S_8$ binds via an S-S bridge site. At +2.00 V, $S_8$ prefers binding to a Cr atop site by a margin of 0.2 eV. The S binding grand energy to the (100) facet has a bias dependence of -0.7 eV/V for the Cr atop site, and -0.1 eV/V for S-S bridge site. The (100) facet is shown in *Figure S6*. Cr-atop binding occurs at all Cr sites. Favorable S-S bridge binding occurs between adjacent S atoms bound to Cr and S atoms bound to Ga, as shown in *Figure S5*. Binding sites of S-S bridge and Cr-atop sites are indicated by red boxes in *Figure S6*.

# SUPPLEMENTAL INFORMATION

Supplemental Information is available from the Wiley Online Library or from the author.

# ACKNOWLEDGEMENTS

This work was supported by the National Science Foundation, Award No. NSF CBET-2323065. The views expressed in this article do not necessarily represent the views of the National Science Foundation or the U.S. Government. We would like to express our sincere gratitude to Dr. Nicholas R. Singstock for his invaluable contributions to the conceptualization and design of this study.

# CONFLICT OF INTEREST

The authors declare no conflict of interest.

# DATA AVAILABILITY STATEMENT

The data that support the findings of this study are available in the supplemental material of this article.

# KEYWORDS

Hybrid cathode, batteries, data-driven materials design, and inverse materials design.